\documentclass[pre,10pt,onecolumn,amsmath,amssymb,aps]{revtex4}
\usepackage{graphicx}
\usepackage{dcolumn}
\usepackage{bm}
\usepackage{color}
\begin{document}

\title{Real-time observation of the isothermal crystallization kinetics in a deeply supercooled liquid}

\author{M. Zanatta}
\altaffiliation[Present affiliation: ]{Dipartimento di Informatica, Universit\`a di Verona, I-37134 Verona, Italy}
\email[E-mail: ]{marco.zanatta@univr.it}
\affiliation{Dipartimento di Fisica e Geologia, Universit\`a di Perugia, I-06123 Perugia, Italy}
\affiliation{ISC-CNR c/o Dipartimento di Fisica, Sapienza Universit\`a di Roma, I-00185 Roma, Italy}
\author{L. Cormier}
\affiliation{Institut de Min\'eralogie, de Physique des Mat\'eriaux, et de Cosmochimie (IMPMC), Sorbonne Universit\'{e}s, UPMC Universit\'e Paris 06, CNRS UMR 7590, Mus\'eum National d'Histoire Naturelle, IRD UMR 206, F-75005 Paris, France}
\author{L. Hennet}
\affiliation{Conditions Extr\^emes et Mat\'eriaux: Haute Temp\'erature et Irradiation, CEMHTI-CNRS, Universit\'e d'Orl\'eans, F-45071 Orl\'eans, France}
\affiliation{Laboratoire Léon Brillouin, CEA-CNRS, CEA Saclay, F-91191 Gif sur Yvette, France}
\author{C. Petrillo}
\affiliation{Dipartimento di Fisica e Geologia, Universit\`a di Perugia, I-06123 Perugia, Italy}
\affiliation{IOM-CNR c/o Dipartimento di Fisica e Geologia, Universit\`a di Perugia, I-06123 Perugia, Italy}
\author{F. Sacchetti}
\affiliation{Dipartimento di Fisica e Geologia, Universit\`a di Perugia, I-06123 Perugia, Italy}
\affiliation{IOM-CNR c/o Dipartimento di Fisica e Geologia, Universit\`a di Perugia, I-06123 Perugia, Italy}

\date{\today}

\begin{abstract}
Below the melting temperature $T_m$ crystals are the stable phase of typical elemental or molecular systems. However, cooling down a liquid below $T_m$, crystallization is anything but inevitable. The liquid can be supercooled, eventually forming a glass below the glass transition temperature $T_g$. Despite their long lifetimes and the presence of strong barriers that produces an apparent stability, supercooled liquids and glasses remain intrinsically metastable state and thermodynamically unstable towards the crystal. Here we investigated the isothermal crystallization kinetics of the prototypical strong glassformer GeO$_2$ in the deep supercooled liquid at 1100~K, about half-way between $T_m$ and $T_g$. The crystallization process has been observed through time-resolved neutron diffraction for about three days. Data show a continuous reorganization of the amorphous structure towards the alpha-quartz phase with the final material composed by crystalline domains plunged into a low-density, residual amorphous matrix. A quantitative analysis of the diffraction patterns allows determining the time evolution of the relative fractions of crystal and amorphous, that was interpreted through an empirical model for the crystallization kinetics. This approach provides a very good description of the experimental data and identifies a predator-prey-like mechanism between crystal and amorphous, where the density variation acts as blocking barrier.
\end{abstract}

\maketitle

\section{Introduction}
\label{Sec:Intro}

From a microscopic point of view, the structure of supercooled liquids and glasses is amorphous. Even though the atomic arrangement shows a local ordering that can extend even beyond first neighbor atoms \cite{ElliottNATURE1991}, it globally retains both the continuous translational and rotational symmetries that are proper to the liquid state. Crystallization breaks up these symmetries that become finite, and the structure rearranges towards a long-range atomic order and a thermodynamically stable phase. The time evolution of this process depends on the system and on its thermodynamic conditions, and it can be considered as the reverse of the medal of the glass transition and the main limit to glass stability \cite{TurnbullCP1969}. In fact, the timescales of cristallization span over a wide interval, ranging from geologically stable systems, e.g. \cite{ZhaoNCom2013}, to nanosecond crystallizing materials, e.g. \cite{GreerNMat2015}. 

The standard approach to describe crystallization is the classic nucleation theory (CNT), whose main ingredients are the processes of nucleation and growth \cite{OxtobyACP1988,KeltonSSP1991}. Within this framework, spontaneous fluctuations in the amorphous system lead to the formation of small crystallites. When one of them exceeds a critical size, i.e. when the nucleation free energy barrier is overcome, the whole system crystallizes. In CNT, the critical nucleus is assumed spherical, and the nucleation barrier is determined by a balance between surface and volume free energy terms. Experiments and simulations in several systems show that this picture seems valid for moderate supercooling, e.g. \cite{GasserSCIENCE2001}, whereas it starts to fail in the deep supercooled region. Here, nucleation and growth processes seem more complex, involving also collective non-diffusive rearrangements \cite{SanzPRL2011,SanzPNAS2014}. Finally, recent papers point out the pivotal role of inhomogeneities in the supercooled liquids as triggers for the nucleation \cite{DargaudAPL2011,RussoSR2012,TanakaEPJ2012}, since the presence of these heterogeneus regions could ease the nucleation process. 

The relevance of crystallization goes well beyond fundamental condensed matter physics and follows from the universality of the glassy state in nature and technology \cite{Kelton}. Typical examples come from geology, where the crystallization of volcanic magmas strongly affects the eruptive style of volcanoes \cite{VetereESR2015}, and from material science, where the rapid crystallizing properties of some chalcogenide glasses are considered to develop fast and reliable permanent memories with nanoseconds switching time, e.g. \cite{GreerNMat2015,LokePNAS2014}.

In this paper we focus on the kinetics of the isothermal crystallization process in the deep supercooled liquid, i.e. for a temperature $T\ll T_m$. In this regime, the viscosity is so high that the system is macroscopically solid and structural rearrangements are still so slow that atoms can be thought as frozen. Nevertheless, the structure of an amorphous solid is never really arrested \cite{RutaNCom2014}, and local non diffusive relaxations can lead to crystallization, as observed in metallic glasses \cite{LeitnerPRB2012}. In general, disordered systems show a hierarchy of excitations down to very small frequency that can contribute to the origin of many of the complex characteristics of the glasses \cite{LinPRX2016}.

As a benchmark system, we choose vitreous germania v-GeO$_2$. Like v-SiO$_2$, v-GeO$_2$ is a covalent oxide glass and a prototype of the strong network forming systems \cite{BoehmerJCP1993}. However, with respect to v-SiO$_2$, v-GeO$_2$ has a rather accessible $T_m=1388$~K and a $T_g\simeq818$~K \cite{BoehmerJCP1993}. From a structural point of view, v-GeO$_2$ glassy network is based on Ge(O$_{1/2}$)$_4$ tetrahedra bound together in a corner-sharing network, which is preserved also in the supercooled liquid \cite{MicolautJPCM2006}. Crystalline GeO$_2$ presents two stable polymorphs at room pressure and temperature: a rutile-like tetragonal structure $(P4_2/mnm)$ \cite{BaurAC1971}, and an $\alpha$-quartz-like structure $(P3_221)$ \cite{SmithAC1964}. The latter is also the stable phase for $T\geq1281$~K. 

Starting from the glass, we approached the supercooled liquid by heating the system up to $T_{exp}=1100$~K. At this temperature, the viscosity is very high, about $10^7$~Pa$\cdot$s, while the diffusion coefficient is about $10^{-18}$~m$^2\cdot$s$^{-1}$, see Ref. \cite{AngellNCG1994}. In this condition, GeO$_2$ is still in a substantially arrested state with dynamical and structural properties very similar to those of the glass \cite{ZanattaJCP2011,CaponiPRB2009}. However, with increasing time the system starts to crystallize, and we observed the kinetics of this process by acquiring a set of static structure factors for about 67~h. Results show the emergence of an $\alpha$-quartz phase in a continuous process that reorganizes the amorphous matrix, eventually leading to a mixed system with a large number of crystalline domains and a small fraction of low-density amorphous regions. The time evolution of both the crystalline and amorphous fraction were interpreted within an empirical model for the crystallization kinetics. This approach provides a very good description of the experimental data and identifies a non linear predator-prey mechanism between crystal and amorphous where the density variation acts as limiting barrier.

\section{Experiments and results}
\label{Sec:Exp}

The static structure factor $S(2\theta)$ of v-GeO$_2$ was firstly measured at room temperature to exclude any appreciable crystallization of the original glass. The temperature was then slowly raised up to 975~K, monitoring the structure to detect any trace of crystallization. Finally, the sample was quickly heated to $T_{exp}=1100$~K with a slope of 20~K/min, and then the 67 hours long isothermal measurement was initiated. In order to properly trace the time evolution of the crystallization, we chose two different acquisition times. During a first period of 16 h, the acquisition time was set to $\Delta t$ = 5 minutes. This is the minimum time to get a good statistics, and short enough to provide detailed view of the beginning of the process. Once the crystallized fraction was clearly visible, the acquisition time was lengthened to $\Delta t=30$~min.

%%%%%%%%%%%%%%%%%%%%%%%%%%%%%%%%%%%%%%%%%%%%%%%%%%%%%%%%%%%%%%%%%%%%%%%%%%%%%%%%%%%%%%%%%%%%%%%%%%%%%%%%%%%%%%%%%%%%%%
\begin{figure}[htb]
	\centering
		\includegraphics[width=0.52\textwidth]{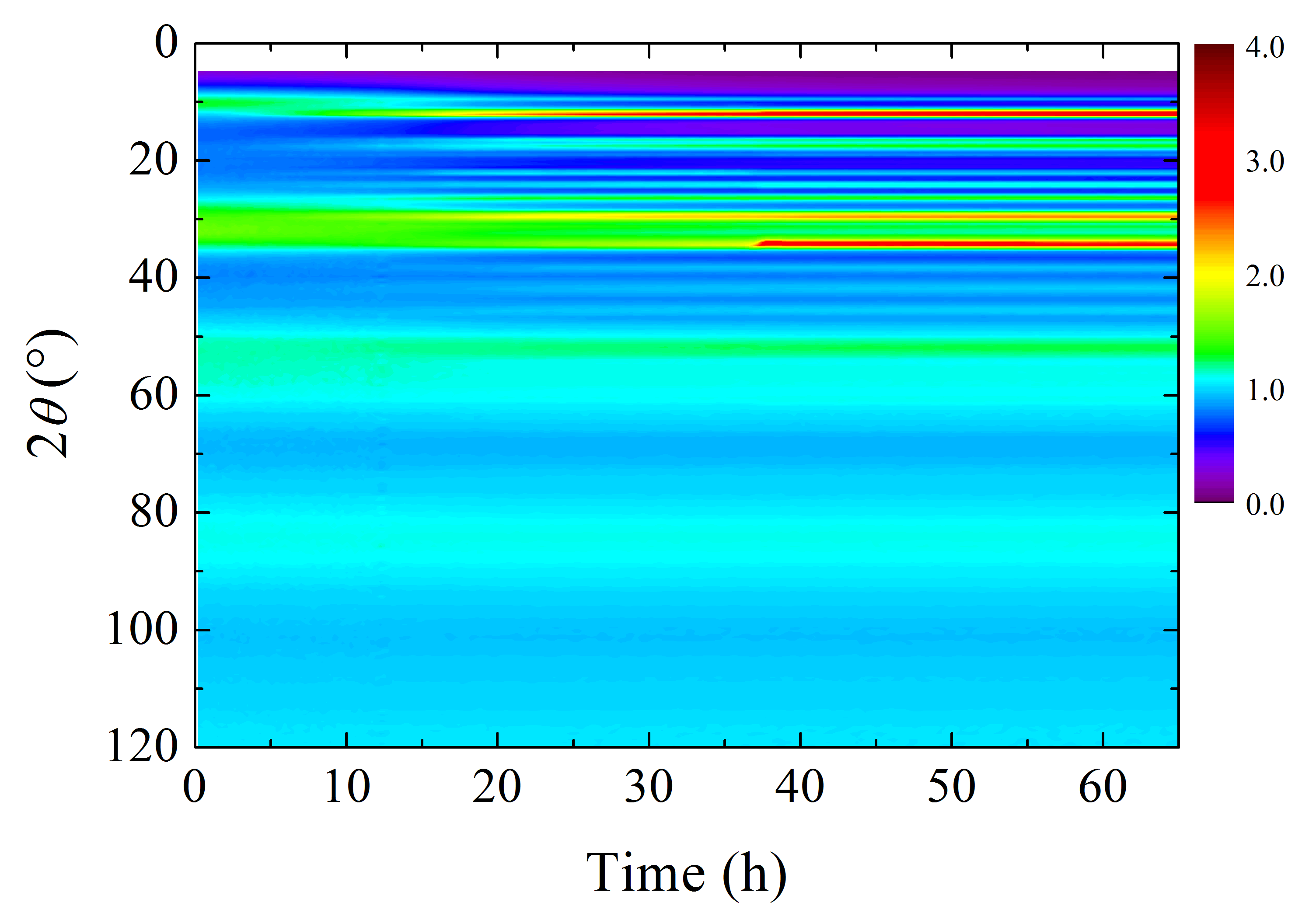}
	\caption{Time evolution of the static structure factor of GeO$_2$ at $T_{exp}=1100$~K. Time increases from left to right. The color map shows the emergence of the crystalline pattern.}
	\label{fig1}
\end{figure}
%%%%%%%%%%%%%%%%%%%%%%%%%%%%%%%%%%%%%%%%%%%%%%%%%%%%%%%%%%%%%%%%%%%%%%%%%%%%%%%%%%%%%%%%%%%%%%%%%%%%%%%%%%%%%%%%%%%%%%

\begin{figure}[htb]
	\centering
		\includegraphics[width=0.48\textwidth]{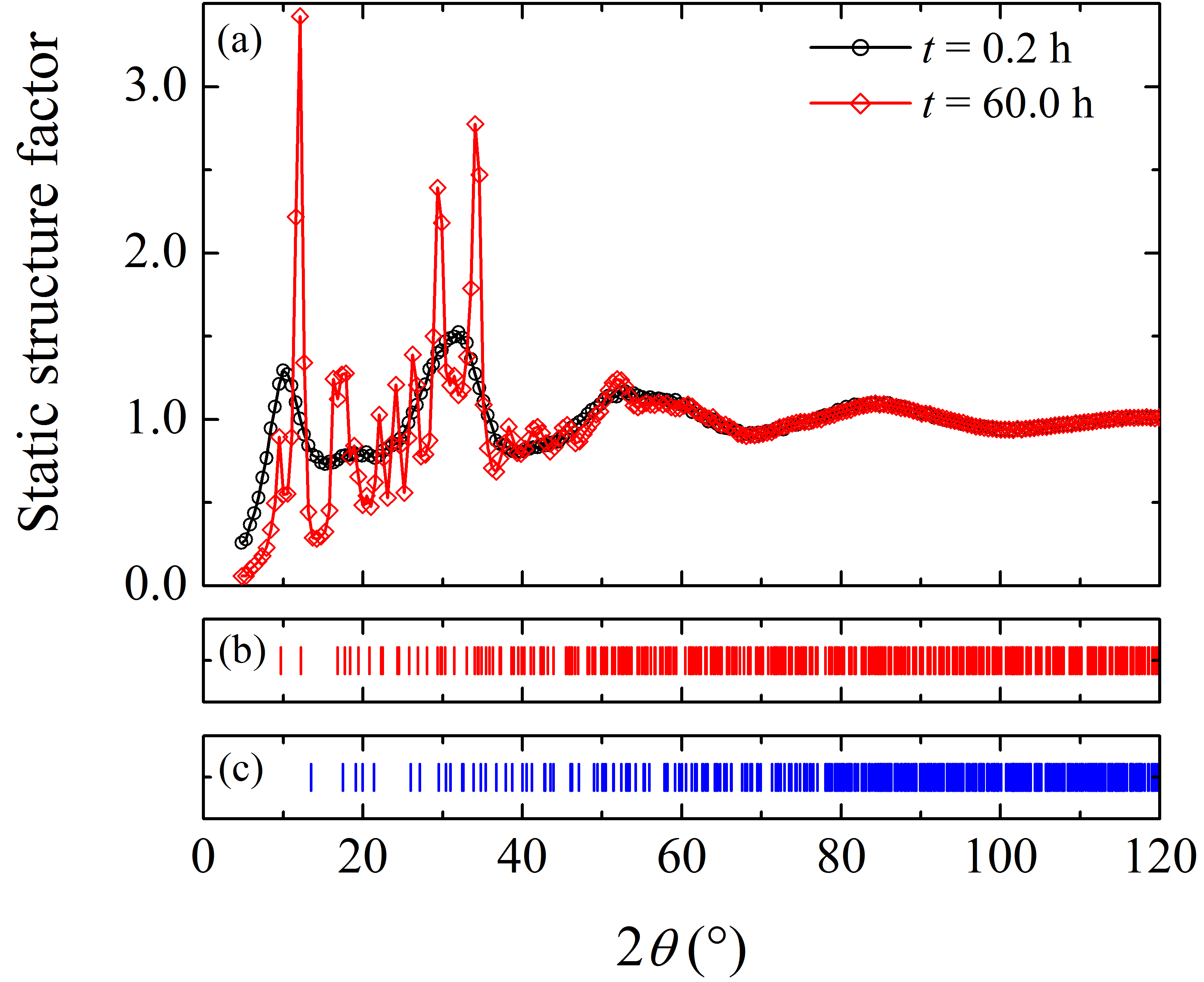}
	\caption{(a) Static structure factor $S(2\theta)$ measured at $t=0.2$~h (black open circles) and at $t=60.0$~h (red diamonds). The solid lines are just connections between experimental points. Calculated Bragg peak positions for GeO$_2$ crystalline polymorphs at room temperature: (b) $\alpha$-quartz-like structure $(P3_221)$, Ref. \cite{SmithAC1964}; (c) rutile-like tetragonal structure $(P4_2/mnm)$, Ref. \cite{BaurAC1971}.}
	\label{fig2}
\end{figure}

The time evolution of the $S(2\theta)$ during the isothermal measurement is reported in Fig. \ref{fig1}. On increasing time, the initially amorphous $S(2\theta)$ shows the growth of a crystalline phase through the appearance of Bragg peaks, clearly visible in the low $2\theta$ part of the diffraction pattern, below about $60^{\circ}$. The peak intensities increase and eventually saturate but the peaks pattern remains the same, without intermediate phases. At high $2\theta$ the Debye-Waller factor reduces the intensity of the Bragg peaks and the $S(2\theta)$ appears substantially unchanged, still keeping glassy-like smooth features. The crystallization process is summarized in Fig.\ref{fig2}(a), that shows a comparison between the first fully amorphous $S(2\theta)$ measured at $T_{exp}$ and one acquired after 60~h. The position of the Bragg peaks observed in the latter is compatible with that of the $\alpha$-quartz \cite{SmithAC1964}, Fig. \ref{fig2}(b), and no traces of rutile-like structure are visible \cite{BaurAC1971}, Fig. \ref{fig2}(c). This also implies that the crystallization process preserves the chemical composition without any appreciable phase separation.  

%%%%%%%%%%%%%%%%%%%%%%%%%%%%%%%%%%%%%%%%%%%%%%%%%%%%%%%%%%%%%%%%%%%%%%%%%%%%%%%%%%%%%%%%%%%%%%%%%%%%%%%%%%%%%%%%%%%%%%
%%%%%%%%%%%%%%%%%%%%%%%%%%%%%%%%%%%%%%%%%%%%%%%%%%%%%%%%%%%%%%%%%%%%%%%%%%%%%%%%%%%%%%%%%%%%%%%%%%%%%%%%%%%%%%%%%%%%%%

\subsection{Determination of crystalline and amorphous fractions}
Assuming that no contributions arise from the crystal-amorphous interfaces, we can write the measured $S(2\theta)$ as the sum of an amorphous term and a crystalline one, namely  
\begin{eqnarray}
	S(2\theta)=S_A(2\theta)+S_C(2\theta).
	\label{Eq:FitSdQ}
\end{eqnarray}
The first term accounts for the amorphous fraction of the material, and  
\begin{eqnarray}
	S_A(2\theta)=A_a S_g(2\theta),
	\label{Eq:FitSdQA}
\end{eqnarray}
where $A_a$ is a parameter and $S_g$ is the static structure factor of the fully amorphous system, obtained by considering the first scans at 1100~K, where no trace of crystallization is visible. Since the static structure factor is the integral over the energy, the crystalline term $S_c(2\theta)$ can be written as the sum of an elastic contribution $S_B(2\theta)$ accounting for the Bragg peaks, and an inelastic part identified as the thermal diffuse scattering (TDS). Consequently,
\begin{eqnarray}
	S_C(2\theta)=S_B(2\theta)+A_c S_T(2\theta),
	\label{Eq:FitSdQC}
\end{eqnarray}
where $A_{c}$ is a parameter. Following Ref. \cite{Warren}, we use a very simple approximation for the TDS contribution $S_T(2\theta)$, i.e.
\begin{eqnarray}
	S_{T}(2\theta)=1-e^{-2W},
	\label{Eq:FitSdQT}
\end{eqnarray}
where $\exp{(-2W)}=\exp{\left(-2B(\frac{\sin\theta}{\lambda})^2\right)}$ is the Debye Waller factor. The parameter $B$ for v-GeO$_2$ at 1100~K  was calculated in harmonic approximation using the vibrational density of states from Ref. \cite{FabianiJCP2008}. Finally, $S_B(2\theta)$ is modeled describing each Bragg peak with a Gaussian, whose position $2\theta_i$ is given by the $\alpha$-quartz structure using the appropriate lattice parameters $a$ and $c$. A preliminary analysis of the Bragg diffraction pattern did not show any appreciable $t$-evolution of the peak width, so we assumed that the peak full width half maximum $\sigma_i\sqrt{2\log2}$ is given by the instrument resolution, see appendix \ref{App:Res}. The Bragg contribution turns out to be:
\begin{eqnarray}
	S_{B}(2\theta)=\sum_i\frac{A_i}{\sigma_i\sqrt{2\pi}}e^{-\frac{1}{2}\left(\frac{2\theta-2\theta_i}{\sigma_i}\right)^2},
	\label{Eq:FitSdQB}
\end{eqnarray}
where $A_i$ is the integrated intensity of the $i$th reflection which is fitted independently for each peak.

%%%%%%%%%%%%%%%%%%%%%%%%%%%%%%%%%%%%%%%%%%%%%%%%%%%%%%%%%%%%%%%%%%%%%%%%%%%%%%%%%%%%%%%%%%%%%%%%%%%%%%%%%%%%%%%%%%%%%%
\begin{figure}[htb]
	\centering
		\includegraphics[width=0.85\textwidth]{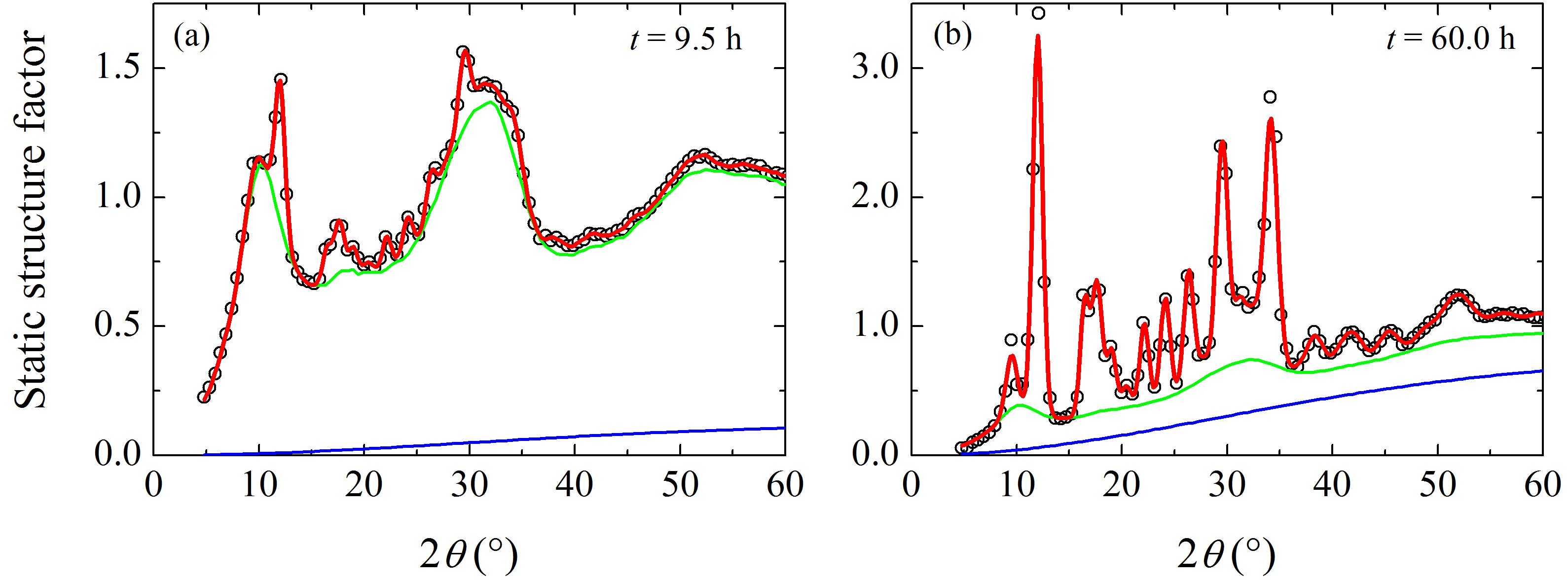}
	\caption{Static structure factor of GeO$_2$ at $T=1100$~K, measured at $t=9.5$~h (a) and $t=60.0$~h (b). The red line is the best fit to the data (black circles) according to Eq. \ref{Eq:FitSdQ}. The solid blue line is $S_{T}$, and the solid green line shows the sum of the amorphous and TDS components, see text.}
	\label{fig3}
\end{figure}
%%%%%%%%%%%%%%%%%%%%%%%%%%%%%%%%%%%%%%%%%%%%%%%%%%%%%%%%%%%%%%%%%%%%%%%%%%%%%%%%%%%%%%%%%%%%%%%%%%%%%%%%%%%%%%%%%%%%%%

The lattice parameters $a$ and $c$ for the $\alpha$-quartz GeO$_2$ at $T_{exp}$ were determined by fitting the most crystallized data with Eq. \ref{Eq:FitSdQ}. This leads to $a(T_{exp})=5.053\pm0.002$~\AA~and $c(T_{exp})=5.66\pm0.04$~\AA, that were then fixed to fit the whole $t$-evolution. As compared to their room temperature counterparts $a(RT)=4.987$~\AA~and $c(RT)=5.652$~\AA~\cite{SmithAC1964}, the high-$T$ values are slightly dilated, and the thermal dilatation seems fairly anisotropic, as it affects $a$ more than $c$.

Eq. \ref{Eq:FitSdQ} provides a good fit to data during the whole observed process. This is visible in Fig. \ref{fig3}, where two examples at two different times are reported: Fig. \ref{fig3}(a) shows the early stage of the crystallization, $t=9.5$~h, while Fig. \ref{fig3}(b) reports the result after 60~h. 

%%%%%%%%%%%%%%%%%%%%%%%%%%%%%%%%%%%%%%%%%%%%%%%%%%%%%%%%%%%%%%%%%%%%%%%%%%%%%%%%%%%%%%%%%%%%%%%%%%%%%%%%%%%%%%%%%%%%%%
\begin{figure}[htb]
	\centering
		\includegraphics[width=0.85\textwidth]{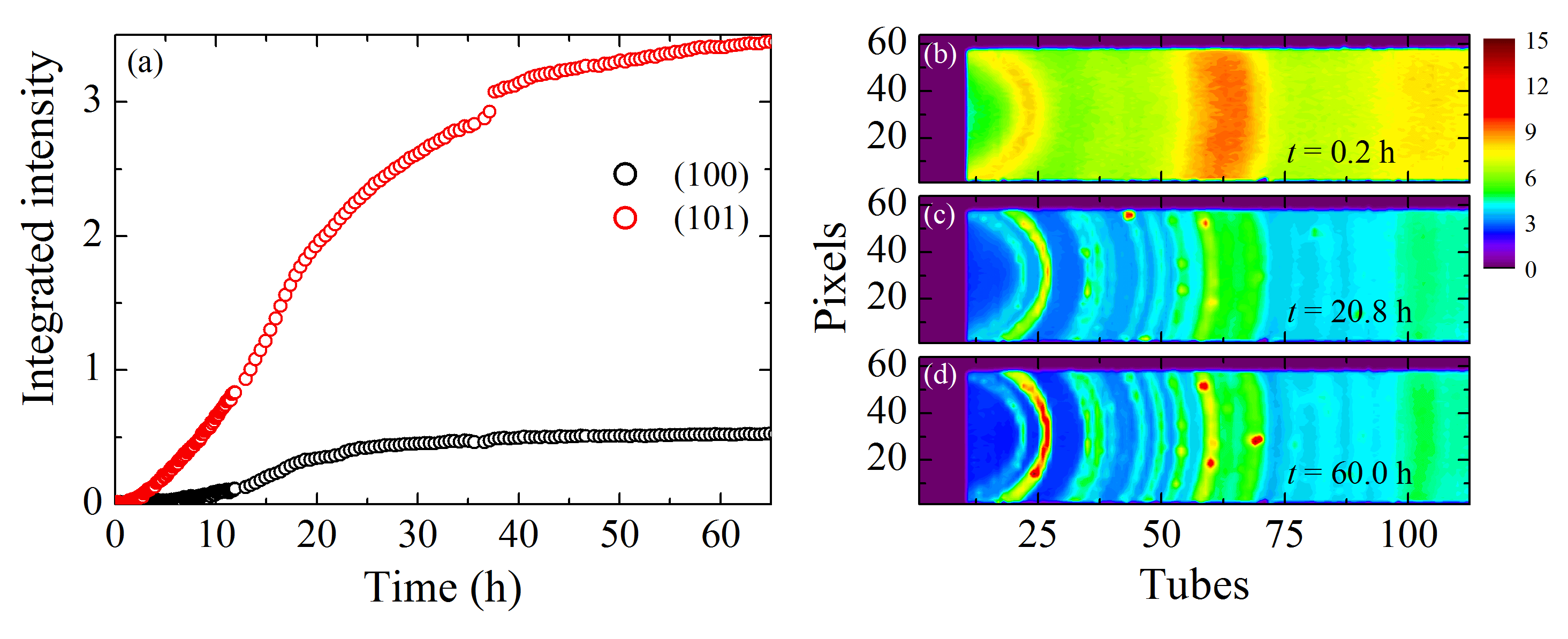}
	\caption{Time evolution of integrated intensity $A_i$ for the first two reflections of the $\alpha$-quartz structure, (100) and (101), see legend. The insets show the intensity on a portion of the detector at three different times during the isothermal process. The color scale of the PSD maps is the same for all the images.}
	\label{fig4}
\end{figure}
%%%%%%%%%%%%%%%%%%%%%%%%%%%%%%%%%%%%%%%%%%%%%%%%%%%%%%%%%%%%%%%%%%%%%%%%%%%%%%%%%%%%%%%%%%%%%%%%%%%%%%%%%%%%%%%%%%%%%%

The time evolution of the integrated intensity of the first two Bragg reflections is reported in Fig. \ref{fig4}. The intensity of the reflections shows a smooth increase as a function of time with an inflection point after about 20~h and a tendency to a long time saturation. However, the (101) reflection displays a step like increase at about $t=36$~h, hardly visible in the (100). To trace the origin of this feature, we can analyze the intensity collected on the PSD detector, which is shown in the inset of Fig. \ref{fig4} at three different times. A typical amorphous pattern, with broad and regular Debye-Scherrer rings, is visible in the upper panel, corresponding to the beginning of the isotherm. After about 20~h, Bragg peaks emerge and the intensity becomes polycrystal-like, with some high intensity spots located on the Debye-Scherrer rings. This suggests that most part of the crystalline phase is basically a polycrystal, i.e. a spherically averaged assembly of small crystalline domains. However, some domains can grow more than the average and, if conveniently oriented, they produce the observed single crystal diffraction, with Bragg spots on the Debye-Scherrer rings. These bigger grains are then modified by the growth of neighboring crystals so the spots can evolve, and eventually disappear when the grains change their orientation.

\begin{figure}[htb]
	\centering
		\includegraphics[width=0.48\textwidth]{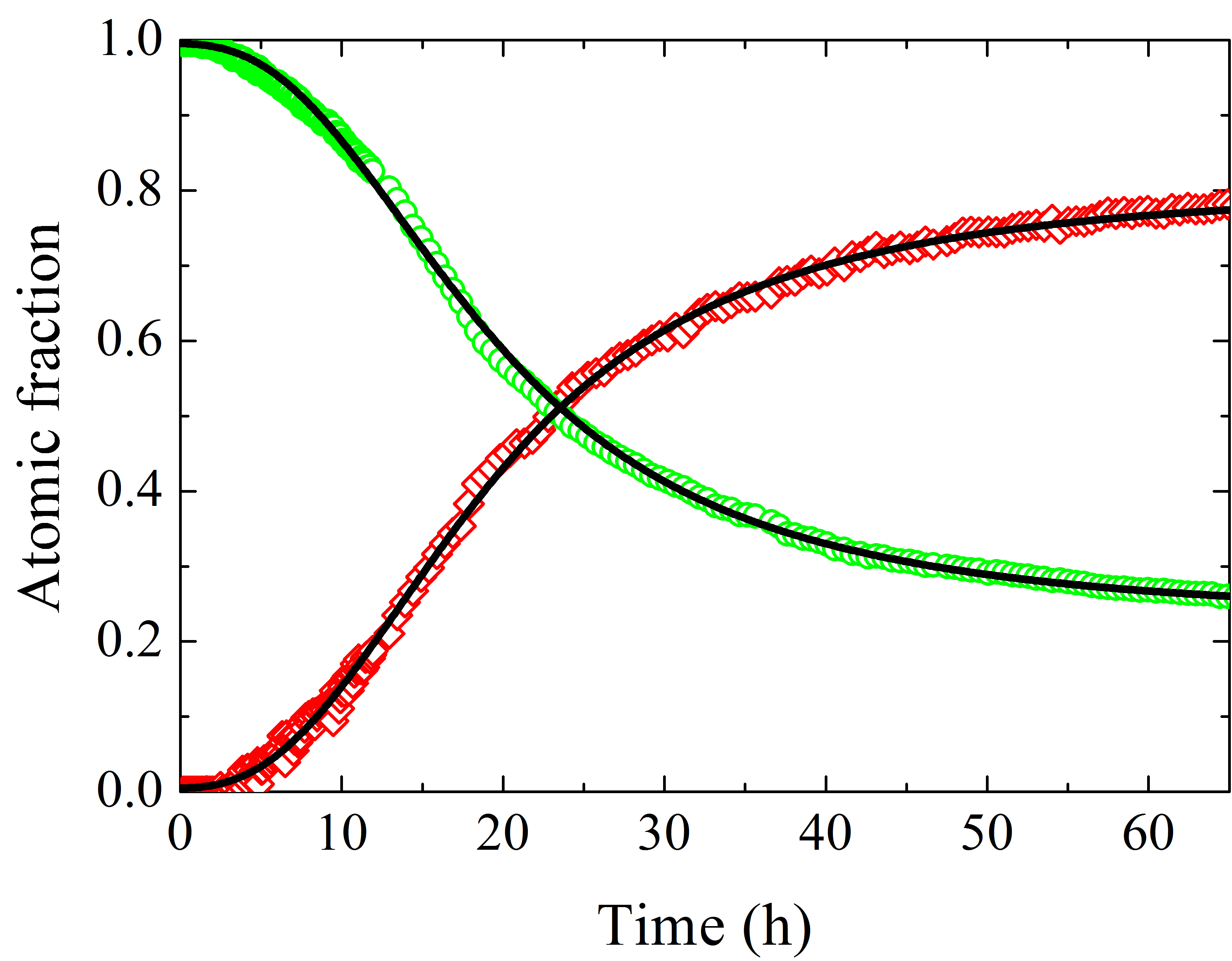}
	\caption{Time evolution of the crystalline and amorphous fractions, $A_c$ and $A_a$, red open diamonds and green open circles respectively. The solid line represents the fit with the model, as described in the text.}
	\label{fig5}
\end{figure}

Considering the above observations, it is clear that the sample cannot be considered as a perfect powder, therefore the fraction of the crystalline phase is not directly accessible. Conversely, we prefer to extract the fraction of atoms in the crystalline and amorphous phase by resorting to the coefficients $A_c$ and $A_a$ of Eqs. \ref{Eq:FitSdQA} and \ref{Eq:FitSdQC}. As a matter of fact, the scattering intensity at high scattering angle is proportional to the number of atoms and the properly normalized $S(2\theta)$ is equal to 1. In this limit, Bragg peaks are suppressed by the Debye-Waller factor and smeared out by the instrument resolution, whereas the TDS and the amorphous static structure factor go to unity. This means that Eq. \ref{Eq:FitSdQ} reduces to $S(2\theta)=A_a+A_c\simeq1$, hence $A_c$ represents the fraction of atoms in the crystalline phase, whereas $A_a$ represents that in the amorphous one. The time evolution of these quantities is reported in Fig. \ref{fig5} and provides an insight into the kinetics of the crystallization process, as well as into the corresponding decrease of the amorphous matrix. In particular, a qualitative analysis of their shapes confirms that crystallization becomes appreciable after 4~h, and then it rapidly develops by subtracting material from the amorphous phase. After about 20~h, the crystallization rate slows down leading to a final material where a 77\% of the atoms is organized in the $\alpha$-quartz structure, whereas the remaining 23\% still shows amorphous features.

%%%%%%%%%%%%%%%%%%%%%%%%%%%%%%%%%%%%%%%%%%%%%%%%%%%%%%%%%%%%%%%%%%%%%%%%%%%%%%%%%%%%%%%%%%%%%%%%%%%%%%%%%%%%%%%%%%%%%%
%%%%%%%%%%%%%%%%%%%%%%%%%%%%%%%%%%%%%%%%%%%%%%%%%%%%%%%%%%%%%%%%%%%%%%%%%%%%%%%%%%%%%%%%%%%%%%%%%%%%%%%%%%%%%%%%%%%%%%

\section{Discussion}
\label{Sec:Discussion}
All the previous experimental observations can be combined to develop an empirical model that describes the kinematic of the crystallization in a deeply supercooled liquid. We thus consider a system of $N_{tot}$ atoms at a temperature $T\ll T_m$, i.e. where diffusion can be safely neglected. At a given time $t$ the system is composed by $N_c$ atoms in the crystalline phase and $N_a$ atoms in the amorphous one. Of course, the relation $N_c(t)+N_a(t)=N_{tot}$ holds at any time. Data suggest the presence of many different crystalline domains evolving in time. According to this, $N_c$ can be written as
\begin{eqnarray}
	N_c(t)=\sum_{i=0}^{N(t)}n_i(t-t_i).
	\label{Eq:Model1}
\end{eqnarray}
where $N(t)$ is the number of crystalline nuclei at a given time $t$, whereas $n_i$ is the number of atoms in the $i$th domain that originates at $t_i$. Considering that the process is almost continuous and both $N(t)\gg1$ and $n_i(t)\gg1$, we can write the sum of Eq. \ref{Eq:Model1} as the time integral 
\begin{eqnarray}
	N_c(t)=\int_{0}^{t}\frac{dN(t')}{dt'}n(t-t')dt'.
	\label{Eq:Model2}
\end{eqnarray}
The function $n(t)$ is zero when $t\le0$, therefore it is convenient to change the integration variable to $\tau=t-t'$, so that Eq. \ref{Eq:Model2} is rewritten as:
\begin{eqnarray}
	N_c(t)=\int_{0}^{t}N'(t-\tau)n(\tau)d\tau,
	\label{Eq:Model3}
\end{eqnarray}
where we use the compact notation $N'(t)=dN(t)/dt$.

Equation \ref{Eq:Model3} describes a process that develops through a nucleation and growth mechanism. The creation of a nucleus is assumed to be a stochastic process that can be triggered by thermal fluctuations in the material and probably eased by even intrinsic heterogeneities \cite{RussoSR2012,TanakaEPJ2012}. Conversely, since diffusion is practically arrested, the growth of each nucleus proceeds only through structural rearrangements involving the interface between the ordered and disordered regions in a self-limiting process. Indeed, each crystalline domain nucleates and grows at expenses of the surrounding amorphous region. However, since the crystal has a higher density than the supercooled liquid, this mechanism creates high-density fully ordered regions that become surrounded by depleted interfaces. In absence of diffusion, this process slows down and stops the growth of the crystalline nuclei. A similar mechanism applies also to nucleation, that becomes less probable in overcrowded and depleted environments.

%%%%%%%%%%%%%%%%%%%%%%%%%%%%%%%%%%%%%%%%%%%%%%%%%%%%%%%%%%%%%%%%%%%%%%%%%%%%%%%%%%%%%%%%%%%%%%%%%%%%%%%%%%%%%%%%%%%%%%
\begin{figure}[bt]
	\centering
		\includegraphics[width=0.48\textwidth]{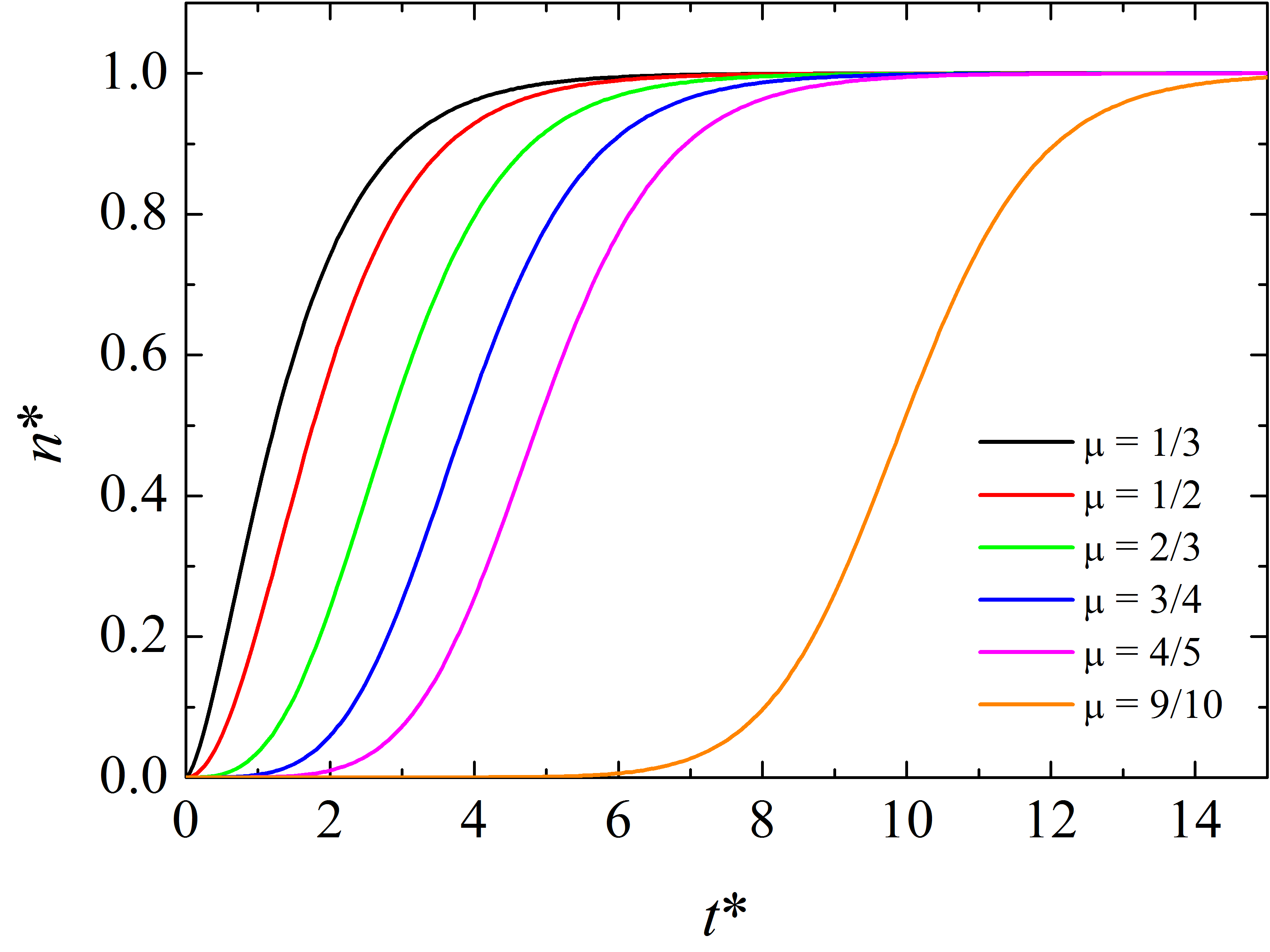}
	\caption{Solution of Eq. \ref{Eq:UniversalGrowth} for different values of the exponent $\mu$.}
	\label{fig6}
\end{figure}
%%%%%%%%%%%%%%%%%%%%%%%%%%%%%%%%%%%%%%%%%%%%%%%%%%%%%%%%%%%%%%%%%%%%%%%%%%%%%%%%%%%%%%%%%%%%%%%%%%%%%%%%%%%%%%%%%%%%%%

According to the previous considerations, the equation for evolution of the number of nuclei $N(t)$ is assumed to be related to a constant rate and can be written as:
\begin{eqnarray}
	\frac{dN}{dt}=\frac{1}{\tau_n}\left[N_m-N\right],
	\label{Eq:Nucleation}
\end{eqnarray}
where $1/\tau_n$ is the rate of the nucleation process and $N_m$ is the maximum number of nuclei.

Conversely, for the growth processes, we can write that
\begin{eqnarray}
	\frac{dn}{dt}=\alpha n^{\mu}(n_m-n),
	\label{Eq:Growth}
\end{eqnarray}
where $\alpha$ is a growth parameter and $n_m$ is the maximum number of atoms in each nucleus. Of course $n_m$ can vary from nucleus to nucleus but, for simplicity, it is assumed to be constant throughout the sample and independent of time . The exponent $\mu$ accounts for the fraction of atoms involved in the process, hence in general $0\leq\mu<1$. Actually, $\mu$ conveys information on the geometry of the nucleus and on the dimensionality $d$ of the process, being its minimum value $1-1/d$. In the case of spherical nuclei, $d=3$ and $\mu$ turns out to be $2/3$. This value seems adequate in the present case. 

By changing the variables and considering $n^*=n/n_m$ and $t^*=t/\tau_d$, with $\tau_d=(\alpha n_m^{\mu})^{-1}$, Eq. \ref{Eq:Growth} turns out to be
\begin{eqnarray}
	\frac{dn^*}{dt^*}=(n^*)^{\mu}(1-n^*),
	\label{Eq:UniversalGrowth}
\end{eqnarray}
This scaled equation does not depend on the actual values of $n_m$ and $\alpha$, and it can be integrated numerically. Typical results for different values of $\mu$ are shown in Fig. \ref{fig6}.

Finally, in a real experimental case, due to the previous thermal history of the sample, we have to consider the possibility that crystallization has already started at $t=0$. This can be easily incorporated into the model by integrating Eq. \ref{Eq:Nucleation} with the condition $N(t=0)=N_0$. Conversely, for the growth, we can use Eq. \ref{Eq:UniversalGrowth} by introducing a fictive time $t_0$ such as
\begin{eqnarray}
	n(t)=n_m n^*\left(\frac{t+t_0}{\tau_d}\right).
	\label{Eq:UniversalGrowth2}
\end{eqnarray}

Within these assumptions, Eq. \ref{Eq:Model2} can be rewritten as
\begin{eqnarray}
	N_c(t)=N_0n_0+\frac{N_m-N_0}{\tau_n}\int^t_0\exp\left(\frac{t'}{\tau_n}\right)n_mn^*\left(\frac{t-t'+t_0}{\tau_d}\right)dt',
	\label{Eq:FitModel}
\end{eqnarray}
where $N_c^0\equiv N_0n_0$ is the initial number crystallized atoms, being $n(t=0)=n_0$. 

Once divided by $N_{tot}$, Eq. \ref{Eq:FitModel} allows for the fit of the crystalline fraction $A_c$ shown in Fig. \ref{fig5}. Moreover, since $N_{tot}=N_A(t)+N_C(t)$, the amorphous fraction $A_a$ can be also analyzed by considering $1-N_c(t)/N_{tot}$. The model has thus five fitting parameters: the initial and the final fraction of atoms in the crystalline phase, $A_c^{(i)}$ and $A_c^{(f)}$, the timescales of nucleation and growth, $\tau_n$ $\tau_d$, and the fictive starting time $t_0$. The parameters can be fitted to $A_c$ and $A_a$ using an overall procedure. The result is reported in Fig. \ref{fig5}. The model provides an accurate and coherent description of both the crystal growth and the amorphous reduction. At the beginning of the process, the system is almost fully amorphous and the crystalline fraction is about $A_c^{(i)}=0.005\pm0.001$ while $A_c^{(f)}=0.77\pm0.01$. However, the presence of a nonzero $t_0$ indicates that the crystallization process is already active, since it starts with a nonzero derivative. The resulting timescales for the nucleation and growth are $\tau_n=16.9\pm0.2$~h and $\tau_d=31.3\pm0.6$~h.

Starting from the density $\rho_m$ of the material resulting from the experiment, we can evaluate the average density of the residual amorphous medium $\rho_a$ at room temperature. The density was measured with a pycnometer and resulted $\rho_m = (3.96\pm0.01)$~g/cm$^3$. Considering $A_a^{(f)}=0.23\pm0.01$ and $A_c^{(f)}=0.77\pm0.01$ with density $\rho_c=4.25$~g/cm$^3$, the average density of the amorphous part turns out to be $\rho_a=(3.25\pm0.03)$~g/cm$^3$, about 91\% of the room temperature glassy value\cite{RieblingJCP1963}, $\rho_g=3.66$~g/cm$^3$. Of course $\rho_a$ is an average value considering both the strongly depleted interfaces and the glassy-like regions. Assuming that the interfaces are fully depleted, we can estimate an upper limit for the true glassy regions as about 89\% of the final amorphous material.

\section{Conclusions}
\label{Sec:Conclusions}
We performed a neutron diffraction study of the isothermal crystallization kinetics in the deep supercooled liquid phase of the strong glassformer GeO$_2$. In the experimental time window, the amorphous system evolves towards a material populated by crystalline $\alpha$-quartz like nuclei plunged into a residual amorphous medium. Due to the difference in density between the two phases, the growth of a crystal causes a rarefaction in the amorphous medium around it. In absence of diffusion, these depleted interfaces act as a barrier for a further growth. The same mechanism involves also the nucleation, that becomes less probable as the nuclei population increases. The density difference between the amorphous and crystalline phases can be considered as a feedback mechanism that controls the growth of the crystal. The so-established predator-prey equilibrium stops further growth at a given crystallite size. In particular, this process introduces a slowing down of the growth function at short time with respect to the simple case, in which diffusion provides material to the nucleating phase. This short time behavior is already evident from the $S(2\theta)$ data of Fig. \ref{fig1}, and it is well described by the model, as shown in Fig. \ref{fig5}. 

The model can be further extended accounting also for the effects of the diffusion. As a matter of fact, even if its contribution seems negligible in the observed timescale, diffusion could give rise to long-time contributions that are expected to become dominant as $T_m$ is approached.

\section*{Materials and methods}
\label{Sec:Methods}
The experiment was carried out at the two axis spectrometer 7C2 \cite{CuelloJPCS2016}. This instrument is located on the hot source of the reactor Orph\'ee at the Laboratoire Leon Brillouin (CEA Saclay, France), and it is optimized for structural studies of liquids and amorphous systems. The Cu (111) monochromator of the instrument was set to obtain an incident neutron wavelength $\lambda=0.724$~\AA. This value, coupled with the high molecular mass of the sample, is short enough to make the inelastic correction negligible. Scattered neutrons were collected on the recently installed position sensitive detector (PSD) based on an assembly of 256 $^3$He tubes. The whole detector covers a scattering angle of 128$^{\circ}$, and each tube has 64 vertical pixels 8~mm height. The calibration of the instrument and the incident wavelength were checked by acquiring the diffraction pattern of a Ni sample. In addition, we measured also a second crystalline standard, a KBr powder, that allowed a thoroughly determination of the resolution function even at low scattering angles, see Supplementary Info.  

Vitreous germania samples were prepared by melt-quenching, starting from Aldrich germanium (IV) oxide crystalline powder (purity higher than 99.998\%). The powder was melted in an alumina crucible at about 1900~K and then quenched in air. Cylindrical specimens with a diameter of 8~mm were core-drilled from the bulk glass, and piled-up to obtain a 50~mm high sample. The sample was loaded in a vanadium cylindrical cell with an outer diameter of 10~mm and 0.5~mm thick walls. The cell was closed with a steel screw cap, which was carefully shielded with boron nitride masks to minimize its scattering contribution. High temperature measurements were done using a vanadium oven under vacuum, $\sim10^{-6}$~mbar. The temperature was monitored by two thermocouples fixed on the body of the sample holder. 

In order to properly evaluate the single scattering intensity of the sample, we collected a set of ancillary measurements including the empty cell, an absorber (a cadmium rod with the same size as the sample container), and the empty beam\cite{PetrilloAC1990,PetrilloAC1992}. The empty cell was also used as vanadium standard to normalize data. The intensity measured on the PSD detector was reduced to $I(2\theta)$ using the program ScRiPT provided by LLB. Starting from these data, the properly normalized static structure factor $S(2\theta)$ was determined via the procedure outlined in Refs. \cite{PetrilloAC1990,PetrilloAC1992}. Monte Carlo simulations were exploited to properly estimate the transmission coefficients and the multiple scattering contribution.

\appendix

\section{Instrument resolution}
\label{App:Res}
The instrument resolution function $\Delta$ was estimated by considering the width of the Bragg peaks of a KBr powder. This material has a $fcc$ structure with a lattice parameter $a=6.598$~\AA~that provides peaks even at relatively low scattering angles. In particular, for $\lambda=0.724$~\AA, the (111) reflection is at $2\theta=10.9^{\circ}$, a value that ensures a reliable determination of the instrument resolution also in the region of the first peaks of the $\alpha$-quartz GeO$_2$.

%%%%%%%%%%%%%%%%%%%%%%%%%%%%%%%%%%%%%%%%%%%%%%%%%%%%%%%%%%%%%%%%%%%%%%%%%%%%%%%%%%%%%%%%%%%%%%%%%%%%%%%%%%%%%%%%%%%%%%
\begin{figure}[htb]
	\centering
		\includegraphics[width=0.48\textwidth]{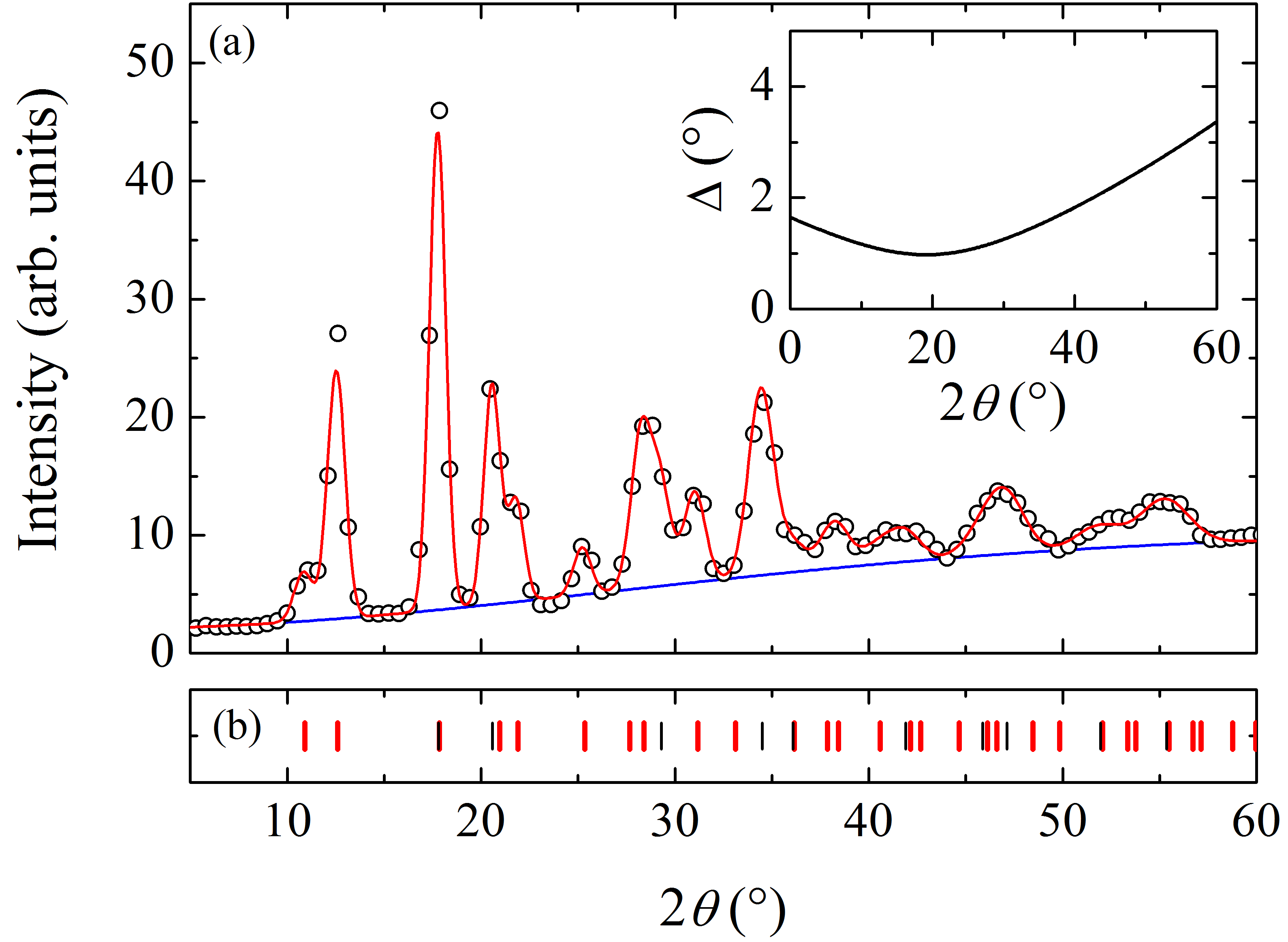}
	\caption{(a) Diffraction pattern measured on KBr powder. Experimental data (black open circles) are fitted by using the model of Eq. \ref{Eq:KBrFit} (red line); the blue line is the sum of $S_T(2\theta)$ and $bkg$. The instrumental resolution $\Delta$ as obtained by the fit is reported in the inset (black line). (b) Positions of the Bragg peaks used in the fit; thick red stiks mark KBr structure whereas the blue ones represent the position of Al peaks due to the container.}
	\label{figA1}
\end{figure}
%%%%%%%%%%%%%%%%%%%%%%%%%%%%%%%%%%%%%%%%%%%%%%%%%%%%%%%%%%%%%%%%%%%%%%%%%%%%%%%%%%%%%%%%%%%%%%%%%%%%%%%%%%%%%%%%%%%%%%

The sample was loaded into an Al cylindrical cell and measured at $T=373$~K to avoid any water contamination of the KBr powder. Data were analyzed according to Refs. \cite{PetrilloAC1990,PetrilloAC1992}. The empty cell contribution was not subtracted and Al Bragg peaks were thus included in the analysis. The so-obtained diffraction pattern is shown in Fig. \ref{figA1}. 

The measured intensity was fitted as the sum of three components
\begin{eqnarray}
	S(2\theta)=S_T(2\theta)+S_B(2\theta)+bkg.
	\label{Eq:KBrFit}
\end{eqnarray}
The first term $S_T(2\theta)$ is the thermal diffuse scattering, $S_B(2\theta)$ represents the Bragg peaks pattern of KBr and Al, whereas $bkg$ is a flat background accounting for incoherent contributions. The TDS contribution was approximated by using Eq. \ref{Eq:FitSdQT}, with $B=(2.33\pm0.09)$~\AA$^2$, see Ref. \cite{ButtAC1976}. Similarly, as in Eq. \ref{Eq:FitSdQB}, $S_B(2\theta)$ is written as a sum of Gaussian functions. The position of the $i$th peak $2\theta_i$ is given by the structure, whereas its area $A_i$ is an independent fit parameter.
Conversely, we assume that the full width at half maximum of each peak is completely given by the instrument resolution $\Delta$. Following Ref. \cite{CagliotiNIM1958} we can write that
\begin{eqnarray}
\Delta=\sqrt{W_2\tan(\theta)^2+W_1\tan(\theta)+W_0},
	\label{Eq:Res}
\end{eqnarray}
where $W_2$, $W_1$ and $W_0$ depend on the instrument collimation and they can be obtained by fitting Eq. \ref{Eq:KBrFit} to the data.

Fig. \ref{figA1} shows the results of the fit on KBr data. The fit was refined by including both KBr and Al peaks, red and black lines in Fig. \ref{figA1}(b) respectively. The instrument resolution determined according Eq. \ref{Eq:Res} is reported in the inset of Fig. \ref{figA1}.

\end{document}